\documentclass[prd,amsmath,amssymb,showpacs,nofootinbib]{revtex4}

\usepackage{graphicx}
\usepackage{dcolumn}
\usepackage{bm}
\usepackage{mathrsfs} 
\usepackage{xspace}

\newcommand{\Z}{\ensuremath{Z}\xspace}
\newcommand{\W}{\ensuremath{W}\xspace}
\newcommand{\bquark}{\ensuremath{b}\xspace}

\newcommand{\mtop}{\ensuremath{m_{t}}\xspace}
\newcommand{\jes}{\ensuremath{S_j}\xspace}
\newcommand{\bjes}{\ensuremath{S_b}\xspace}
\newcommand{\jetsmear}{\ensuremath{R}\xspace}
\newcommand{\mtopreco}{\ensuremath{m_{t}^{\rm reco}}\xspace}
\newcommand{\jesreco}{\ensuremath{S_j^{\rm reco}}\xspace}
\newcommand{\bjesreco}{\ensuremath{S_b^{\rm reco}}\xspace}

\newcommand{\mtoprecohad}{\ensuremath{m_{t,\,\rm had}^{\rm reco}}\xspace}
\newcommand{\mtoprecolep}{\ensuremath{m_{t,\,\rm lep}^{\rm reco}}\xspace}
\newcommand{\ddeltamtoprecolephaddbjes}{\ensuremath{{M^{\rm reco}}}\xspace}

\newcommand{\mw}{\ensuremath{m_\W}\xspace}
\newcommand{\mwraw}{\ensuremath{m_\W^{\rm raw}}\xspace}
\newcommand{\El}{\ensuremath{{E_\ell}}\xspace}
\newcommand{\plz}{\ensuremath{{p_\ell^z}}\xspace}

\newcommand{\ptlvec}{\ensuremath{{\vec{p}_\ell^{\,T}}}\xspace}

\newcommand{\pnuz}{\ensuremath{{p_\nu^z}}\xspace}

\newcommand{\ptnuvec}{\ensuremath{{\vec{p}_\nu^{\,T}}}\xspace}

\newcommand{\ptmiss}{\ensuremath{p \kern-0.6em\slash_{T}}\xspace}
\newcommand{\ptmissvec}{\ensuremath{\vec{p} \kern-0.6em\slash_{\rm T}}\xspace}

\newcommand{\dzero}{D0\xspace}

\newcommand{\bbbar}{\ensuremath{b\bar{b}}\xspace}
\newcommand{\ttbar}{\ensuremath{t\bar{t}}\xspace}

\newcommand{\ljets}{\ensuremath{\ell}+\rm jets\xspace}

\newcommand{\alpgen}{{\sc alpgen}\xspace}

\newcommand{\ipb}{\ensuremath{\rm pb^{-1}}\xspace}
\newcommand{\MeV}{\ensuremath{\mathrm{Me\kern-0.1em V}}\xspace}
\newcommand{\GeV}{\ensuremath{\mathrm{Ge\kern-0.1em V}}\xspace}
\newcommand{\TeV}{\ensuremath{\mathrm{Te\kern-0.1em V}}\xspace}

\newcommand{\DeltaR}{\ensuremath{\Delta {\cal R}}\xspace}

\begin{document}
\title{Independent measurement of the top quark mass and
the light- and bottom-jet energy scales at hadron colliders}

\author{                                                                      
F.~Fiedler
}
\affiliation{                                                                 
\centerline{Ludwig-Maximilians-Universit{\"a}t M{\"u}nchen,            
  M{\"u}nchen, Germany}                                      
}                                                                             

\date{June 12, 2007}
           
\begin{abstract}

A method for the simultaneous determination of the energy scales for
\bquark-quark jets and light jets, the
jet energy resolution, and the top quark mass at
hadron colliders is presented.  
The method exploits the unique
kinematics of events with top-antitop pair production, where one of
the top quarks involves a leptonic and one a hadronic \W boson decay.
The paper shows a feasibility study of how this simultaneous measurement
can be performed at
the upcoming LHC experiments ATLAS and CMS.
\end{abstract}
\pacs{14.65.Ha, 12.15.Ff, 29.90.+r}

\maketitle

\section{Introduction}
\label{intro.sec}
Precise knowledge of the energy scale for 
\bquark-quark jet reconstruction will be crucial for many measurements at the 
ATLAS and CMS experiments at the 
upcoming LHC collider at CERN, including decays of the top quark, Higgs boson,
and supersymmetric particles.
The determination of the absolute energy scale for calorimeter jets 
has however proved to be a challenging task for the CDF and \dzero experiments 
at the Fermilab Tevatron collider.
For light jets, 
this scale can be determined from collider data
using events containing an isolated energetic photon
balanced by a jet~\cite{bib-jes}.
In contrast to the calibration of charged-lepton reconstruction,
massive resonances decaying to a jet-jet final state cannot readily
be used for jet energy calibration; for example, the reconstruction of 
$\Z\to\bbbar$ decays is difficult because of the large
backgrounds.
The only exception is the reconstruction of hadronic \W decays in
events containing a top-antitop pair, where one of the top 
quark decays (charge conjugate modes are implicitly included
throughout this paper) involves a hadronic, and the other a leptonic
\W decay (``\ljets \ttbar events'').
Based on such events, 
both CDF and \dzero have performed simultaneous measurements of the 
top quark mass and the calorimeter jet energy scale \jes for
light-quark jets~\cite{bib-insitu}.
No separate determination of the \bquark-quark jet energy scale is 
available so far, and in these measurements, the uncertainty on the 
relative difference between the energy scales for light-quark and 
\bquark-quark jets leads to one of the largest remaining systematic
uncertainties on the top quark mass.

In this paper, a novel measurement technique is presented that allows
a simultaneous determination of the top quark mass \mtop, light-quark jet
energy scale \jes, relative light- to \bquark-quark jet energy scale
\bjes, and the 
jet energy resolution \jetsmear from \ljets \ttbar events recorded at a hadron 
collider.
The energy scale \jes is defined like for example in~\cite{bib-me},
i.e.\ a scale larger than $1.0$ corresponds to a calorimeter response
larger than unity, which means that 
the reconstructed jet energy is too large on average.
The factor \jes is applied to all jets, and an additional factor
\bjes is introduced to describe differences between light-quark and
\bquark-quark jets.
The method to determine all four parameters
relies on the fact that the dependence on \bjes is in general
different for the two reconstructed
invariant top quark masses per event.
This is because the transverse momentum (relative to the beam) 
of the neutrino from the
leptonic \W decay is reconstructed by imposing transverse momentum
balance of the event, which means that it depends on \bjes if the magnitude of the
vector sum of the two \bquark-quark transverse 
momenta is large, while there
is no such dependence for the hadronic \W decay.
Events with small vector sum of the two \bquark-quark transverse
momenta cannot contribute independent information on \mtop and \bjes.

In the following it is assumed that calorimeter noise subtraction and 
the full calorimeter calibration up to a constant scale (which may
be different for light and \bquark-quark jets)
have been performed before the technique described here is applied.
Noise studies can be performed with events recorded with minimal
trigger requirements; the pre-calibration can be based on the
transverse momentum balance in dijet events.
In broad terms, the method to determine the top quark mass and 
absolute jet energy scales then comprises the following steps:
A sample of candidate \ljets \ttbar events that contain independent
information on \mtop and \bjes is selected.
To measure the four parameters \mtop, \jes, \bjes, and \jetsmear,
{\em estimators} are calculated for each selected event.
The number of estimators does not have to be equal to the number of 
parameters to be determined; in the analysis described here, the
four parameters are measured using three estimators
\mtopreco, \jesreco, and \bjesreco which are chosen so that they are
related with \mtop, \jes, and \bjes.
Functions are derived to describe
the expected estimator distributions ({\em templates})
for any given set of assumed 
values of the parameters
\mtop, \jes, \bjes, and \jetsmear.
The measured estimator distributions in the data are then compared to
these fitted templates, and the 
\mtop, \jes, \bjes, and \jetsmear values and their uncertainties and 
correlations are determined in a fit.
To validate the procedure, the analysis is performed on simulated
{\em pseudo-experiments}.
To be precise, since the method is based on a fit to simulated
templates, it determines the {\em ratio} of 
energy scales in data and simulated events.

The method should be applicable to both the LHC and the Tevatron experiments;
as a concrete example, the experimental situation at the LHC
(proton-proton collisions at a center-of-mass energy of $14\,\TeV$)
is considered in the following.
Signal \ttbar \ljets events are generated at parton level and smeared
according to a parameterized detector resolution.
No gluon radiation and no backgrounds are included, and it is assumed
that the two 
\bquark-quark jets are identified unambiguously, i.e., the efficiency
$\epsilon_b$
for \bquark-quark jet identification 
may be smaller than unity, but no light jets are
misidentified.
The aim of this paper is thus to establish the feasibility of the 
calibration technique for such an idealized case; conditions that are
closer to reality and systematic uncertainties will be addressed in a
later publication.
In general, it cannot be assumed that the \bquark-quark jet energy
scale measurement is unbiased by the restriction to those \bquark-quark jets that
are identified.
However, this is not a drawback as most applications, i.e.\ measurements 
for which a
\bquark-quark jet energy scale determination is needed, will also use
identified \bquark jets.
Residual biases due to differences in event topologies between
top-antitop events and other event classes will have to be studied with 
simulated events, but can be considered a second-order effect.

The paper is organized as follows:
Section~\ref{reconstruction.sec} contains a general discussion of the
kinematic reconstruction of \ljets \ttbar events used in the
measurement technique presented here.
The generation of simulated events and the event selection are
described in Section~\ref{simulation.sec}, and results of 
a toy-Monte Carlo study
of the reconstruction method, based on realistic parameters for the 
resolution of the LHC detectors, are discussed in Section~\ref{toymc.sec}.
Section~\ref{outlook.sec} summarizes the findings and 
gives an outlook and conclusion.

\section{Kinematic Event Reconstruction}
\label{reconstruction.sec}
This section describes the event-by-event computation of the 
estimators \mtopreco, \jesreco, and \bjesreco.
It should be noted that these calculations can be performed for 
any \ljets \ttbar candidate event and that the validity of this 
section is not restricted to the simplifying assumptions made in the
generation of events used later in this paper.
While the information in templates obtained with more
realistic conditions may be somewhat diluted, the general arguments
made in this section will still hold.

The signature of a \ljets \ttbar event in the detector is the 
presence of four energetic jets (two \bquark-quark jets from the top and 
antitop, and two light-quark jets from the hadronic \W decay), 
an isolated energetic charged lepton (only electrons and muons are 
considered here, as $\tau$ leptons decay close to the interaction 
point leading to additional complications), and missing transverse
momentum (with respect to a balanced event)
due to the undetected neutrino.
There may be additional jets from gluon radiation, 
however such jets often have lower transverse
energy than the \ttbar decay products.
Such jets are not considered here,
but a minimum jet transverse energy is always required in the
experimental event
selection (such a cut is also included in the study described in 
Section~\ref{simulation.sec}).

When the masses of the six \ttbar decay products are assumed to be 
known, there are 18 unknowns
from the 6 final-state particle 3-momenta per event, and in addition
3 unknowns corresponding to the estimators \mtopreco, \jesreco, and \bjesreco
to be calculated on an event-by-event basis.
There are 17 measurements in each event corresponding to the
five 3-momenta of the four jets and the charged lepton and to
the two components of the missing transverse momentum.
(In fact, the missing transverse momentum is a derived quantity,
calculated from the vector sum of the momenta of all other
final-state particles and other ``unclustered energy'' in the event
that is not assigned to final-state jets.
By imposing transverse momentum balance, the neutrino 
transverse momentum can be identified with the
missing transverse momentum,
but its reconstructed value is not 
independent of the reconstructed jet and charged-lepton momenta.)
In addition to the above 17 measurements, 4 constraints can be applied
in each event, corresponding to the invariant masses of the two \W
bosons and two top quarks.
Given that there are as many
measurements as unknowns, the event kinematics can be solved and the
three estimators \mtopreco, \jesreco, and \bjesreco calculated.

The correct assignment of jets to final-state quarks is not known a
priori, but it is assumed that the two light-quark jets can be
assigned unambiguously to the hadronic \W decay.
The known mass \mw of the hadronically decaying \W boson yields
information on the energy scale \jes for light-quark jets.
It is assumed that the scale factor is the same for both jets, such 
that the estimator \jesreco is calculated as
\begin{equation}
  \label{jesreco.eqn}
    \jesreco = \frac{ \mwraw }
                    { \mw }
  \,
\end{equation}
where \mwraw is the mass obtained from the reconstructed jet energies
and momenta.
For the following discussion, the measured energies and momenta of all
jets are divided by the estimator \jesreco obtained in the same event, 
and the reconstructed
missing transverse momentum is adjusted accordingly.

There are two possibilities for combining 
the two light-quark jets with one of the \bquark-quark jets to 
reconstruct a top-quark decay.
To proceed, one can either select one combination (as done in 
Section~\ref{simulation.sec}), or consider both of them, possibly
together with a suitable relative weight.
For a given combination, a scan over \bjesreco values is performed.
Given an assumed value of \bjesreco, the reconstructed \bquark-quark
jet energies and momenta are scaled accordingly, and
the missing transverse momentum is adjusted and taken as 
transverse momentum of the neutrino from the leptonic \W
decay.
The longitudinal neutrino momentum \pnuz is then obtained from
the known mass \mw of the leptonically decaying \W as
\begin{eqnarray}
  \nonumber
    \mw^2
  & = &
      2 E_\ell \sqrt{ \left(\ptnuvec\right)^2 + \left(\pnuz\right)^2 }
    - 2 \left( \ptlvec \cdot \ptnuvec + \plz \pnuz \right)
  \\
  \label{pnuz.eqn}
  \Leftrightarrow
    \pnuz
  & = &
    \plz \frac{ \frac{1}{2}\mw^2 + \ptlvec \cdot \ptnuvec }
              { \El^2 - \plz^2 }
    \pm \sqrt{   \left( \plz \frac{ \frac{1}{2}\mw + \ptlvec \cdot \ptnuvec }
                                  { \El^2 - \plz^2 } \right)^2
               + \frac{ \left( \frac{1}{2}\mw^2 + \ptlvec \cdot \ptnuvec \right)^2 - \El^2 \left(\ptnuvec\right)^2 }
                      { \El^2 - \plz^2 }
             }
  \, .
\end{eqnarray}
If the solutions are real, one obtains one value for the reconstructed
mass \mtoprecolep
of the top quark with the leptonic \W decay for each \pnuz
solution.
The reconstructed mass \mtoprecohad of the top quark with the hadronic
\W decay is also computed for the given \bjes value.
If one finds $\mtoprecolep=\mtoprecohad$, then this top quark mass
and the corresponding \bjesreco value can be taken as estimator values
for the event.
There are up to four solutions per event: two assignments of
\bquark-quark jets and two solutions to Equation~(\ref{pnuz.eqn}).

\section{Simulation and Event Selection}
\label{simulation.sec}
The kinematic event reconstruction and the measurement of 
the four quantities \mtop, \jes, \bjes, and \jetsmear have been tested
with the experimental resolution of the ATLAS and CMS experiments
in mind.
The \alpgen~\cite{bib-alpgen} program has been used to generate
\ljets \ttbar events 
in proton-proton collisions at a center-of-mass energy of $14\,\TeV$
for true top quark masses between 160 and $190\,\GeV$
in steps of $5\,\GeV$.
To simulate the effect of jet reconstruction in the experiment, the
energies of the final-state quarks have been smeared according to 
a Gaussian resolution whose width $\sigma(E)$ is set to
\begin{equation}
  \label{jetresolution.eqn}
    \sigma(E)
  = 
    \jetsmear \sqrt{E}
\end{equation}
with values of \jetsmear between $0.8$ and
$1.2\,\sqrt{\GeV}$ varied in steps of $0.1\,\sqrt{\GeV}$
(this corresponds to a resolution $\sigma(E)/E$ between $80\%/\sqrt{E}$ and
$120\%/\sqrt{E}$ with $E$ in units of \GeV).
The jet energy resolution of the ATLAS and CMS experiments is expected
to be of the 
same order of magnitude~\cite{bib-TDRs}.
Finally, all jet energies and momenta are multiplied by a factor
\jes between $0.8$ and $1.2$ in steps of $0.1$, and independently, 
another multiplicative scale factor of \bjes between $0.8$ and $1.2$
in steps of $0.1$ is applied to \bquark-quark jet energies and
momenta.

A preselection of \ljets \ttbar candidate events is first applied:
\newcounter{preselection.listcnt}
\begin{list}{\bf (P\arabic{preselection.listcnt})}{\usecounter{preselection.listcnt}
                        \setlength{\itemsep}{1ex}
                        \setlength{\parsep}{0ex}
                        \setlength{\topsep}{1ex}}
\setcounter{preselection.listcnt}{0}
\item
A minimum charged-lepton transverse energy of $20\,\GeV$ is required.
\item
The pseudorapidity $\eta$ of the charged lepton and all jets
must be between $-2.5$ and $2.5$.
\item
The angular separation 
$\DeltaR = \sqrt{\left(\Delta\eta\right)^2 + \left(\Delta\phi\right)^2}$ 
between the charged lepton and any jet
must be larger than $0.4$, and the same must be true for any jet-jet
pair.
\item
After smearing and energy scaling, the transverse energies of all jets must be
larger than $30\,\GeV$.
\end{list}
The estimator \jesreco is calculated from the reconstructed mass of
the hadronically decaying \W as in
Equation~(\ref{jesreco.eqn}), and all reconstructed jet energies and
momenta are scaled by $1/\jesreco$.
Events in which the jet assignment is unambiguous and that 
contain independent information on the top quark mass and
\bquark-quark jet energy scale are then selected by these criteria:
\newcounter{selection.listcnt}
\begin{list}{\bf (S\arabic{selection.listcnt})}{\usecounter{selection.listcnt}
                        \setlength{\itemsep}{1ex}
                        \setlength{\parsep}{0ex}
                        \setlength{\topsep}{1ex}}
\setcounter{selection.listcnt}{0}
\item
To unambiguously assign the two \bquark-quark jets, it is required
that exactly one of the 
\bquark-quark jets have an invariant mass together with the
light-quark jet pair between 150 and $200\,\GeV$.
In the following, this \bquark-quark jet is assumed to come from 
the decay of the top quark with the hadronically decaying \W.
For a sample generated with $\mtop=175\,\GeV$, $\jes=\bjes=1$, and
$\jetsmear=1\sqrt{\GeV}$,
$74\%$ of the events that pass the preselection fulfill this
criterion.
\item
As explained in Section~\ref{intro.sec}, events with small magnitude
of the vector sum of \bquark-quark jet transverse momenta cannot
yield independent information on the top quark mass and \bquark-quark
jet energy scale.
Therefore, this magnitude is required to be larger than $50\,\GeV$.
$80\%$ of the events that pass cut (S1) are retained.
\item
Events are only retained if exactly one solution to
Equation~(\ref{pnuz.eqn}) is found with $0.5<\bjesreco<2.0$ and 
$150\,\GeV<\mtopreco<200\,\GeV$.
Thus a choice between several possible estimator values is avoided.
Of the events that pass criterion (S2), 38\%, 45\%, and 17\%
have 0, 1, and 2 such solutions, respectively.  More than 2
solutions per event do not occur.
Thus, $45\%$ of the events that pass selection criterion (S2) also
pass this cut.
\item
Finally, the quantity 
$\ddeltamtoprecolephaddbjes 
:= \frac{\partial\left(\mtoprecolep - \mtoprecohad\right)}
        {\partial\bjesreco}
$
is obtained during the scan of \bjesreco values and
taken as a measure of whether an event contains
independent information on the top quark mass and \bquark-quark jet
energy scale.
The \ddeltamtoprecolephaddbjes distribution in all events that
pass cut (S3) is shown in Figure~\ref{dmtopdifference.fig}.
\begin{figure}[tb]
\begin{center}
\includegraphics[width=0.49\textwidth]{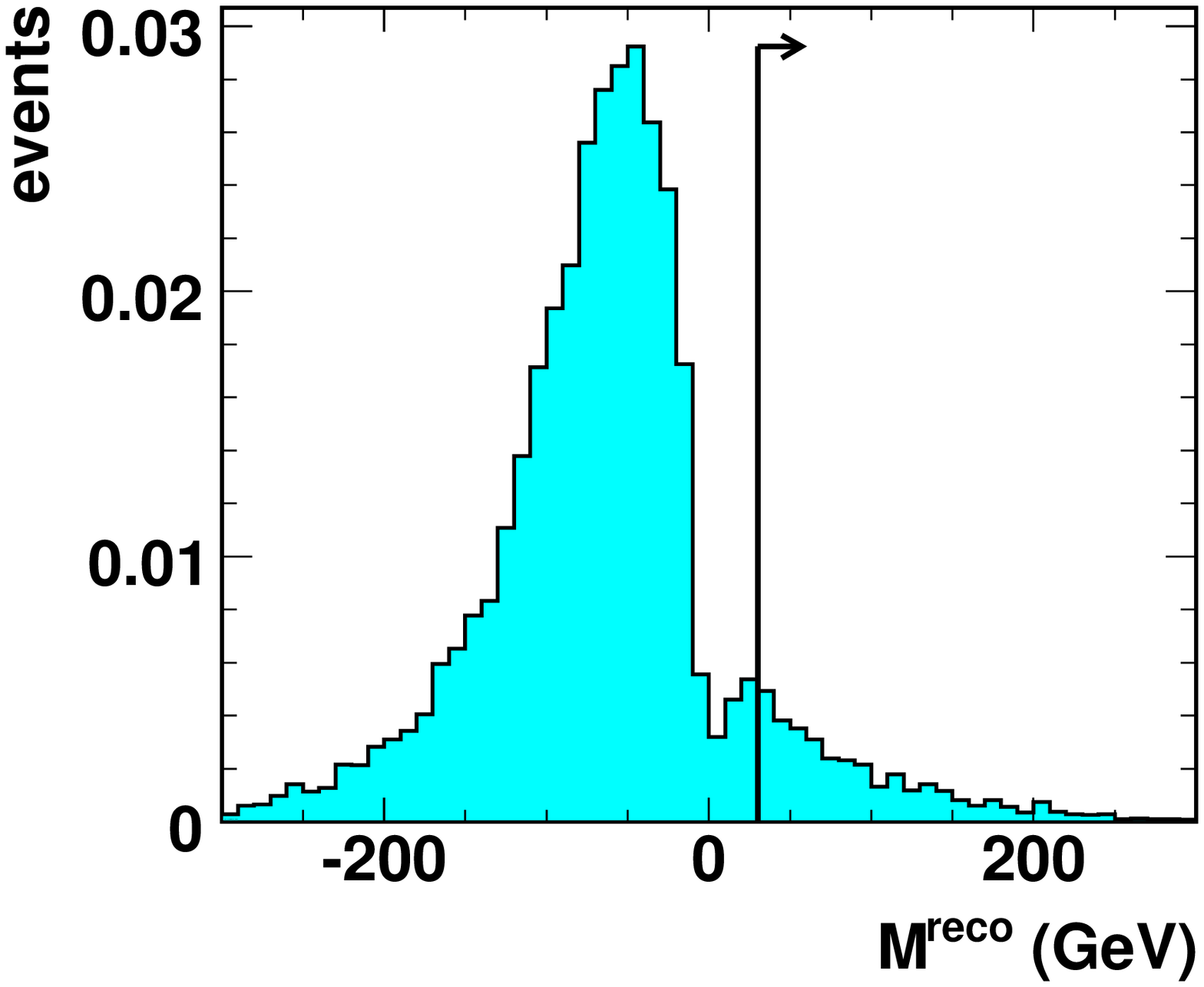}
\caption{\label{dmtopdifference.fig}
  For simulated events with $\mtop=175\,\GeV$, $\jes=\bjes=1$, and
  $\jetsmear=1\sqrt{\GeV}$, the distribution of the 
  quantity \ddeltamtoprecolephaddbjes for events passing
  selection cut (S3).  The selection applied in (S4) is indicated.
  The dip at $\ddeltamtoprecolephaddbjes=0$ is due to the fact that
  such events cannot yield a solution for $\mtoprecolep=\mtoprecohad$.}
\end{center}
\end{figure}
In events with small $\left|\ddeltamtoprecolephaddbjes\right|$
the value of \bjesreco has a large uncertainty, and thus events
with $\left|\ddeltamtoprecolephaddbjes\right|<30\,\GeV$ are rejected.
The \mtopreco and \bjesreco templates for
events with $\ddeltamtoprecolephaddbjes<-30\,\GeV$ have a degraded
resolution compared with the distributions for
$\ddeltamtoprecolephaddbjes>+30\,\GeV$, and only the latter events
are retained.
This criterion thus selects events in which the value of \mtoprecolep 
decreases much less rapidly than \mtoprecohad or even increases with
\bjesreco. 
A fraction of $11\%$ of events passes this final cut, yielding
an overall efficiency of $2.9\%$ after the preselection.
\end{list}

\begin{figure}[ht]
\begin{center}
\includegraphics[width=0.49\textwidth]{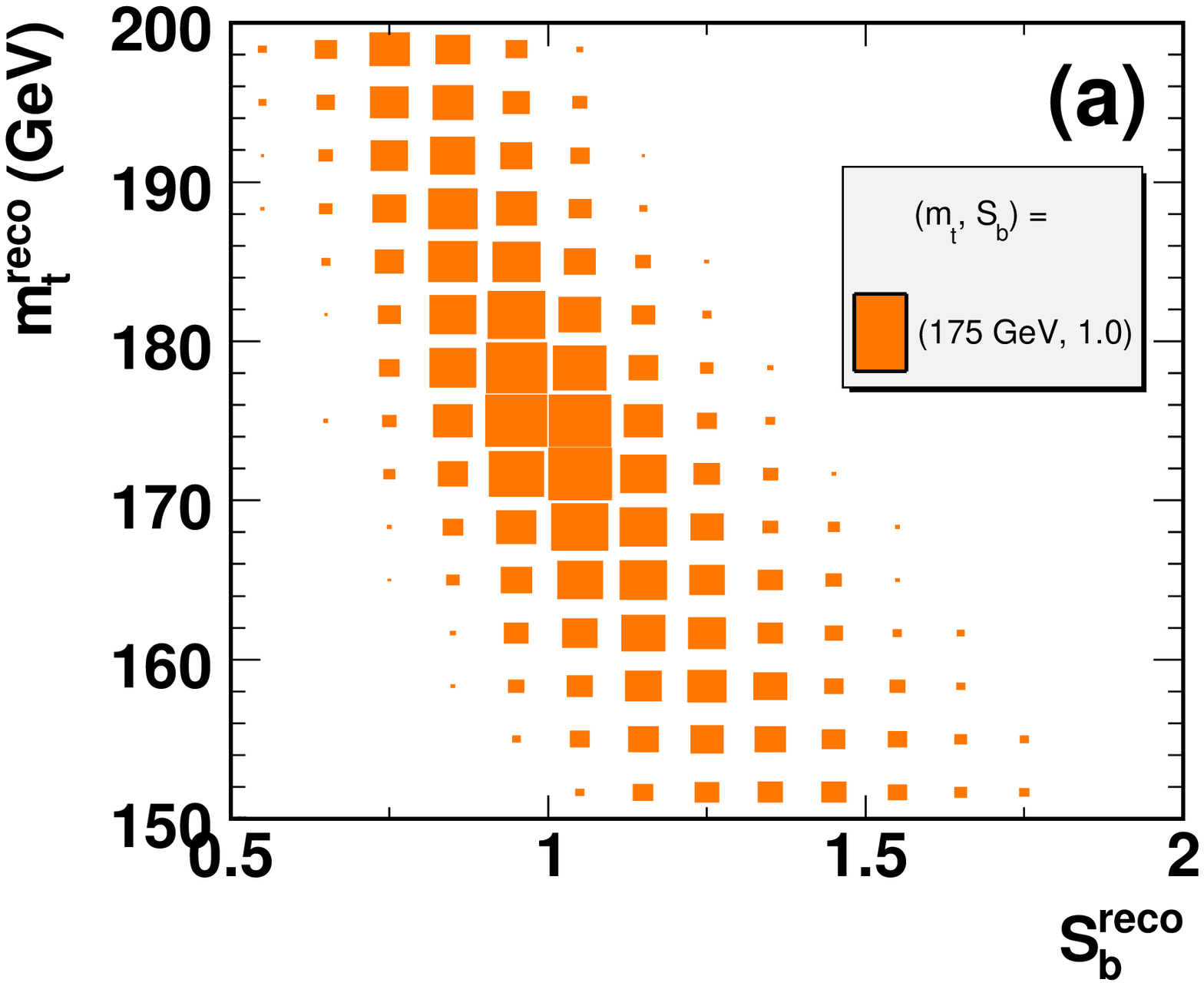}
\hspace{-0.03\textwidth}
\includegraphics[width=0.49\textwidth]{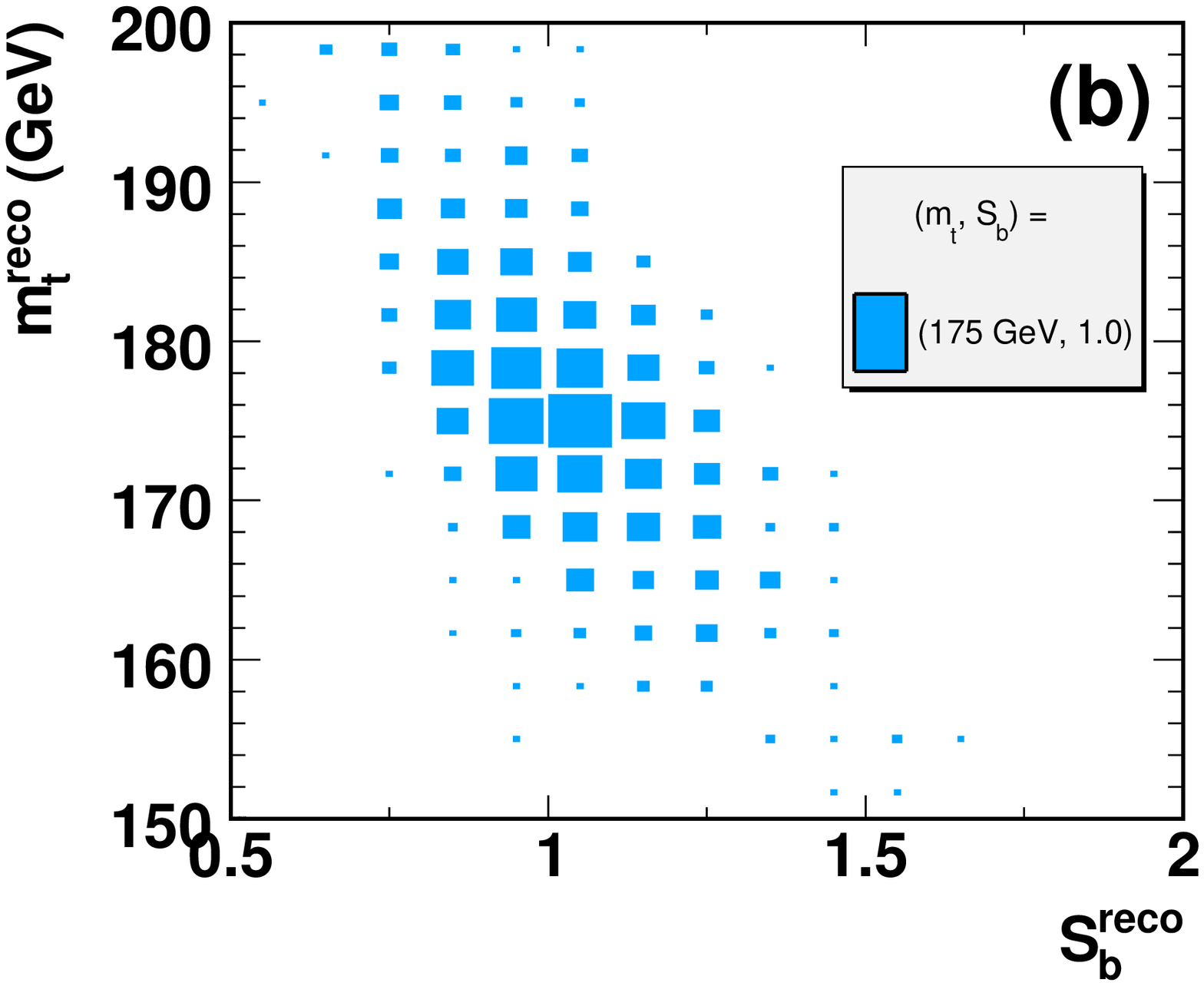}\\
\includegraphics[width=0.49\textwidth]{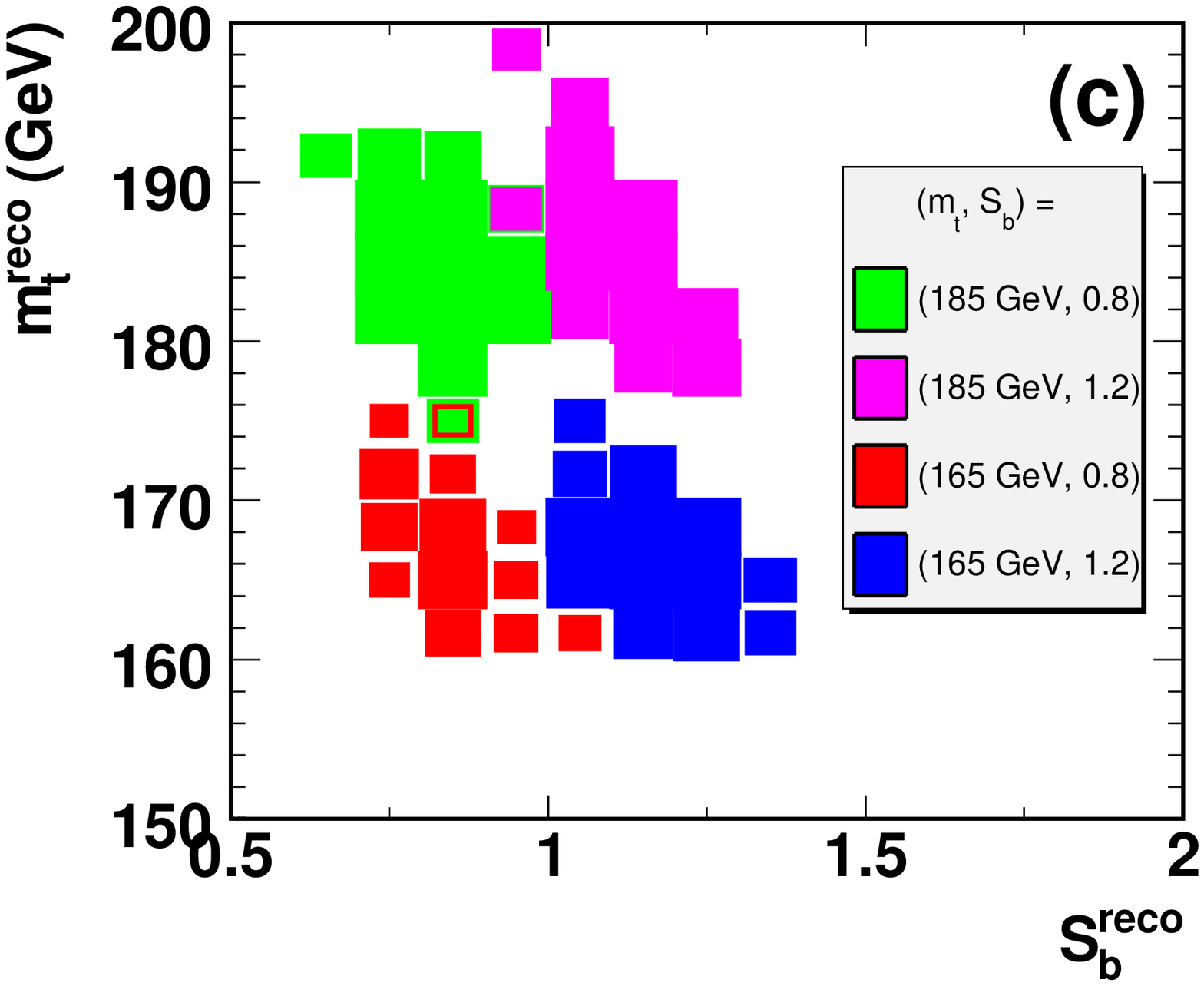}
\caption{\label{mtopbjesestimators.fig}
  For simulated events with $\mtop=175\,\GeV$, $\jes=\bjes=1$, and
  $\jetsmear=1\sqrt{\GeV}$, the two-dimensional \mtopreco vs.\
  \bjesreco estimator distribution
  is shown if (a) cuts (S2) and (S4) are not applied and (b)
  the full event selection is applied.  
  In (c), two-dimensional estimator distributions are shown with
  separate color codes for 
  (\mtop,\,\bjes) input values of 
  ($185\,\GeV$, $0.8$): green, 
  ($185\,\GeV$, $1.2$): magenta,
  ($165\,\GeV$, $0.8$): red, and
  ($165\,\GeV$, $1.2$): blue.
  The input values of
  $\jes=1$ and $\jetsmear=1\sqrt{\GeV}$ have been kept.  
  Here, only those bins are shown that contain at least
  $2\%$ of the events.  The
  reconstructed estimator distributions cluster around
  the input \mtop and \bjes values.
  In all plots, the size of the squares is proportional
  to the number of entries.}
\end{center}
\end{figure}

The effect of criteria (S2) and (S4) is illustrated in 
Figure~\ref{mtopbjesestimators.fig}: Plot (a) shows the 
\mtopreco vs.\ \bjesreco estimator distribution
for a sample generated with $\mtop=175\,\GeV$, $\jes=\bjes=1$, and
$\jetsmear=1\sqrt{\GeV}$ when all selection cuts except (S2) and (S4)
are applied; plot (b) shows events that pass the full event selection.
It is evident how the \mtopreco and \bjesreco resolution is 
superior in those events that pass all selection cuts; the 
correlation is reduced from $-0.76$ to $-0.62$.
Plot (c) shows the \mtopreco vs.\ \bjesreco estimator distributions
for events passing the full selection
for various input values; it can be seen that independent
information on \mtop and \bjes is contained in the events.

It should be noted that the present analysis is intended as a
conceptual study and a proof of principle.
Consequently, no optimisation of the cut values in the 
individual selection criteria
has been performed.

\begin{figure}[p]
\begin{center}
\includegraphics[width=0.49\textwidth]{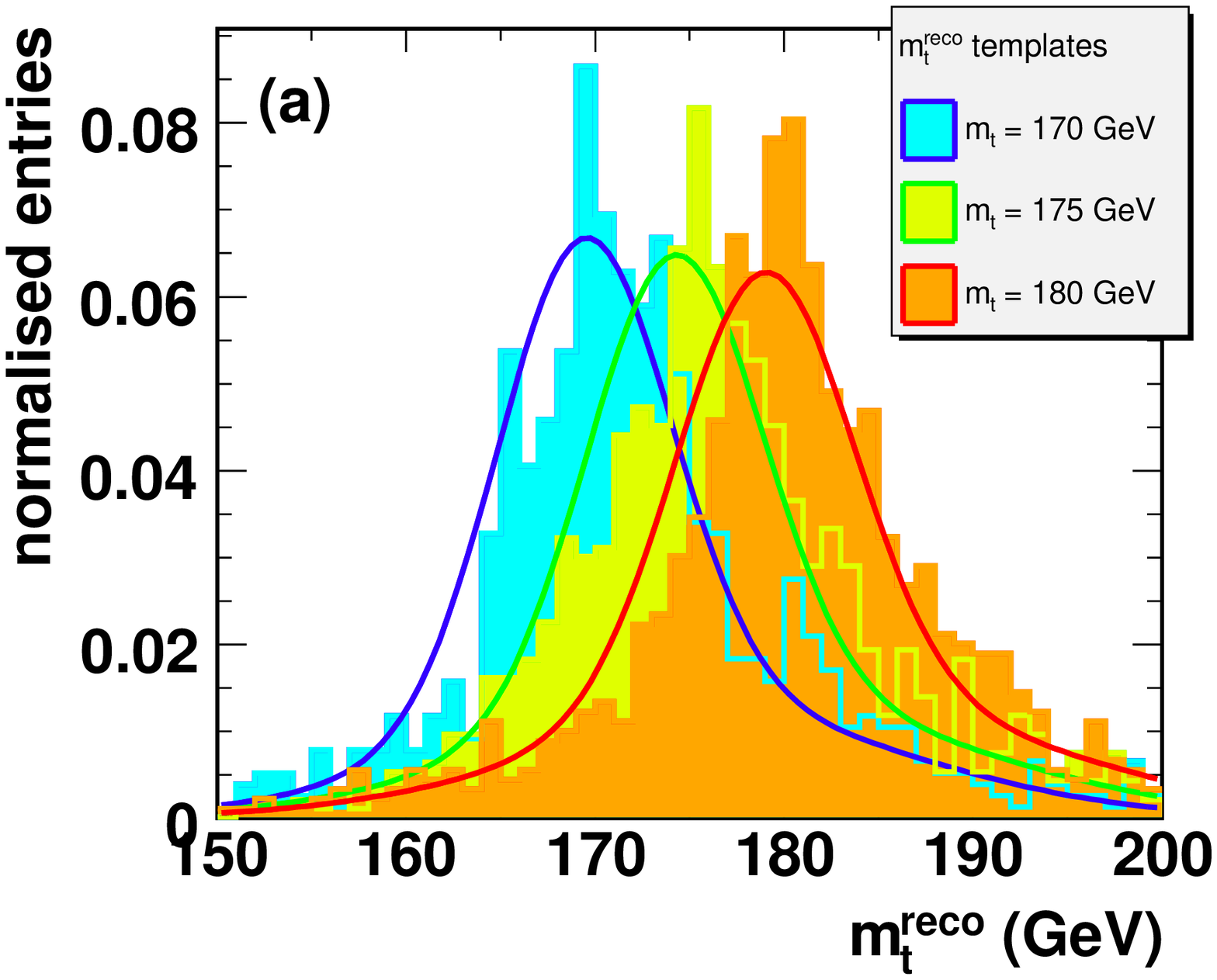}
\hspace{-0.03\textwidth}
\includegraphics[width=0.49\textwidth]{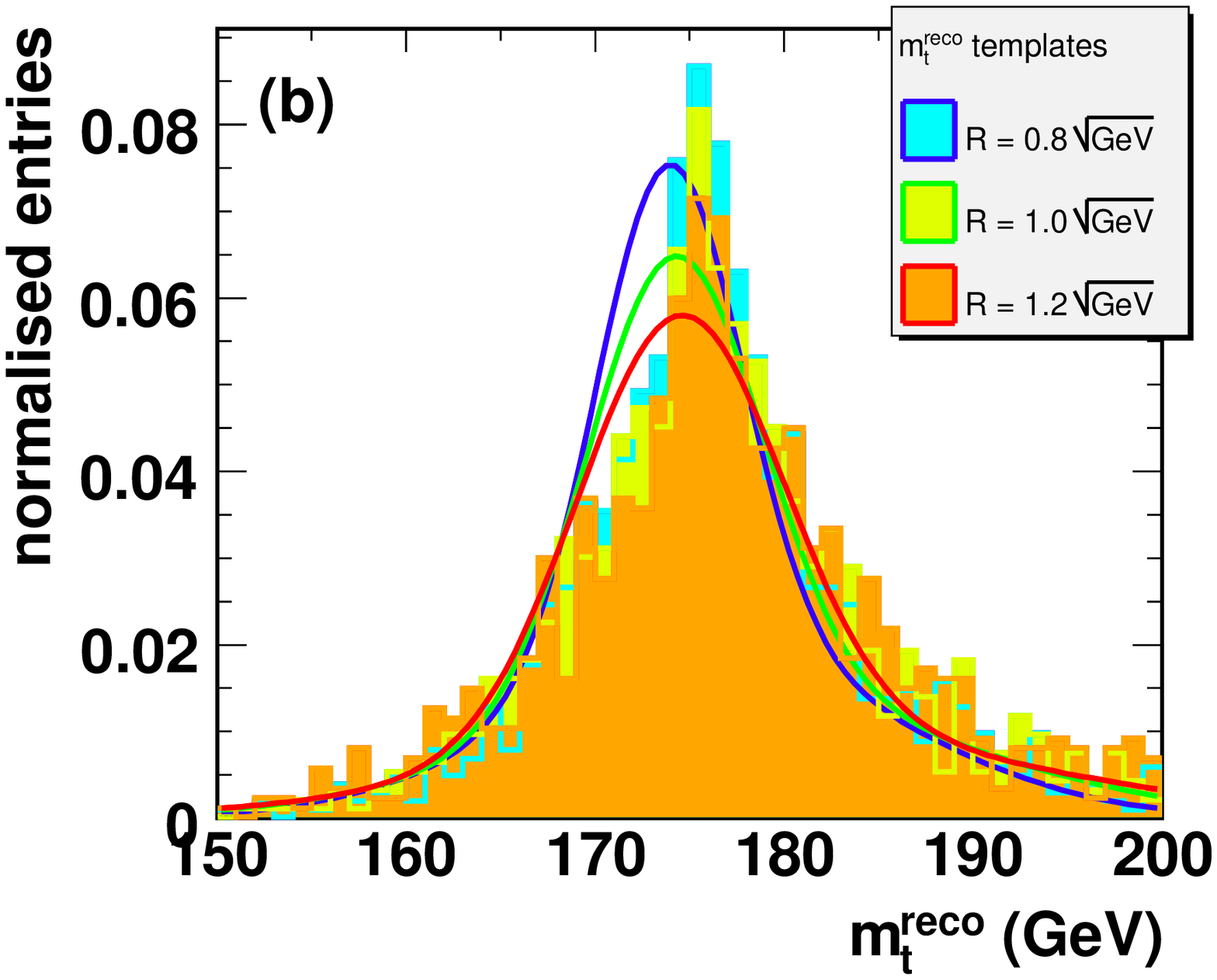}\vspace{-1ex}
\includegraphics[width=0.49\textwidth]{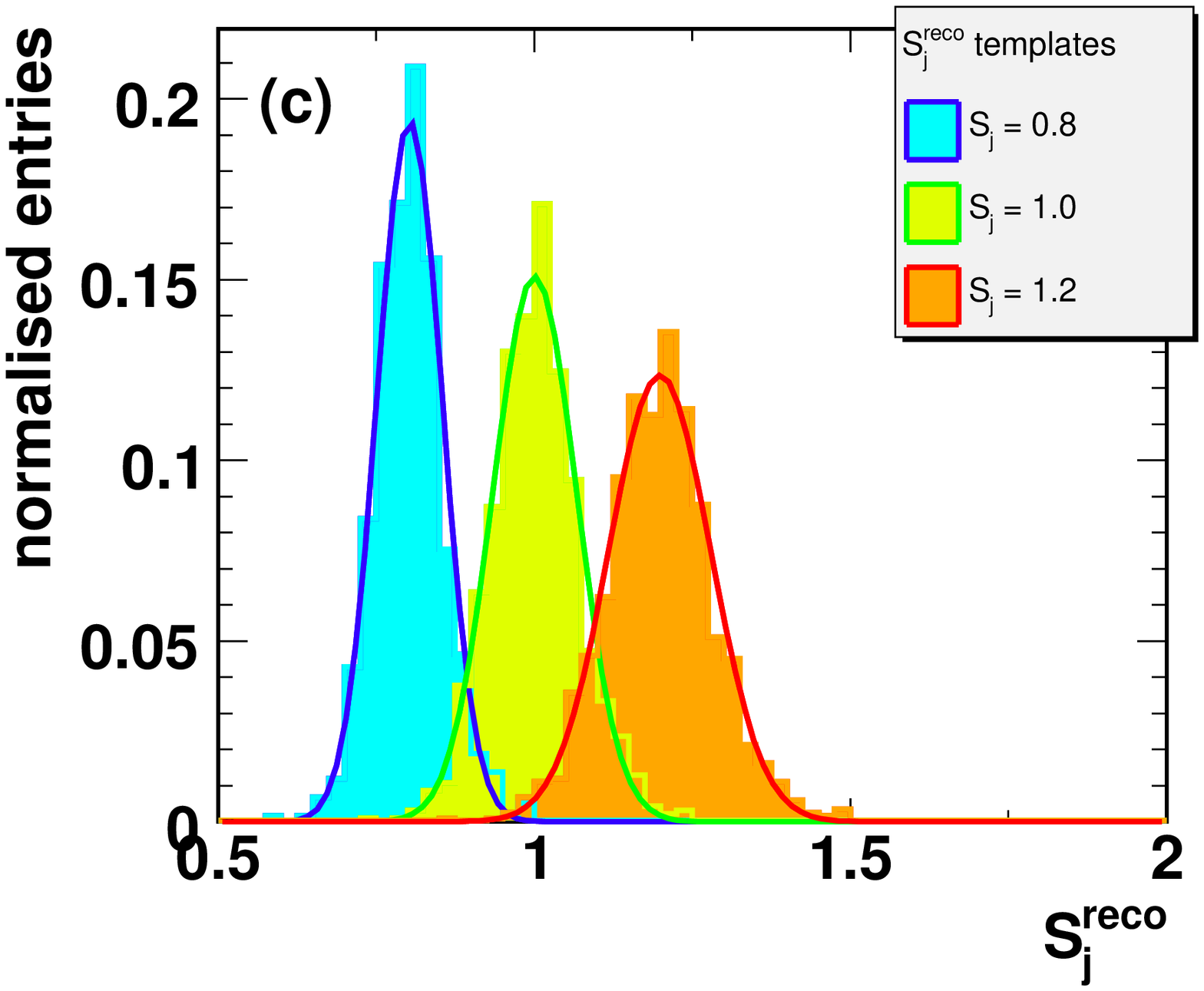}
\hspace{-0.03\textwidth}
\includegraphics[width=0.49\textwidth]{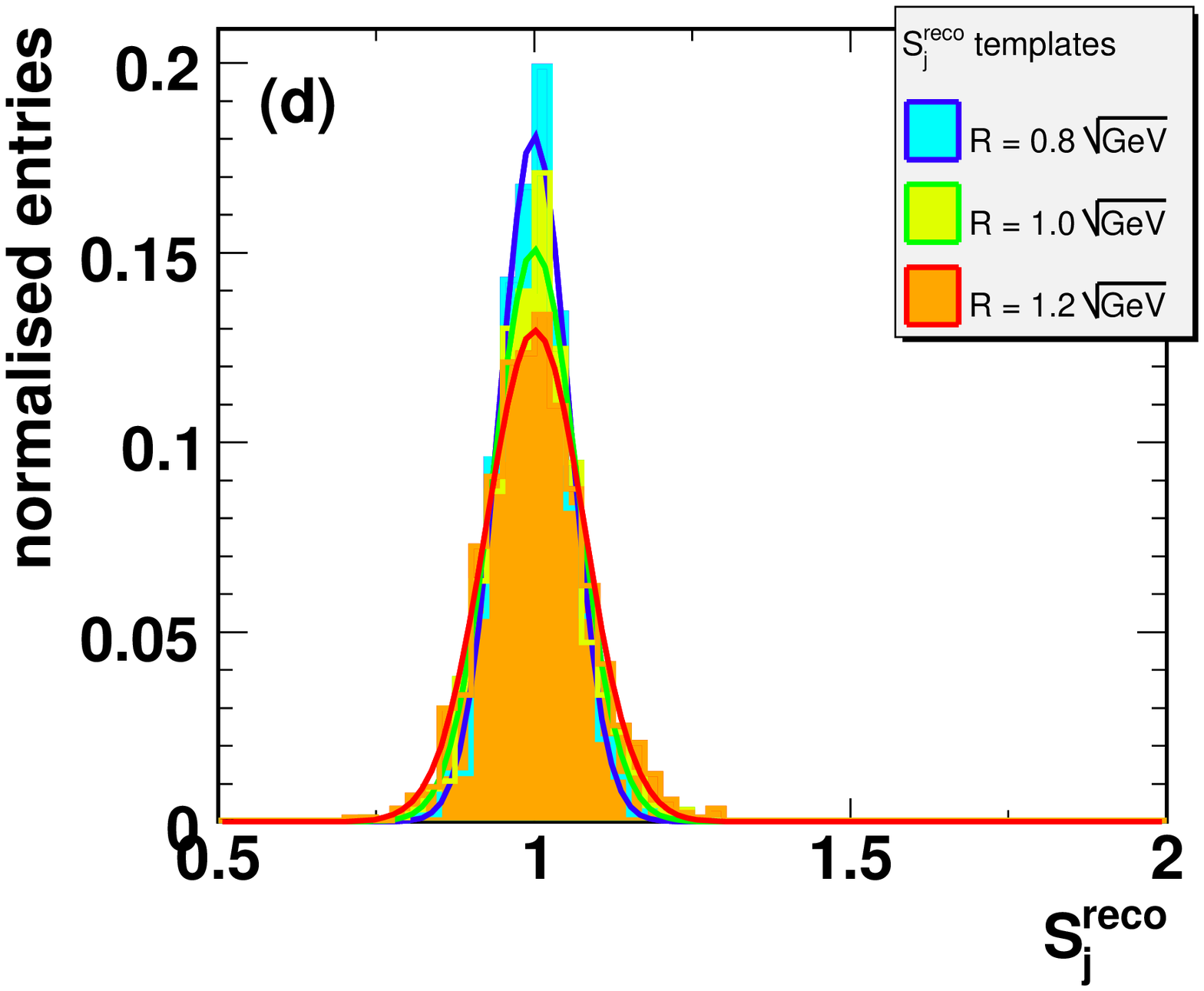}\vspace{-1ex}
\includegraphics[width=0.49\textwidth]{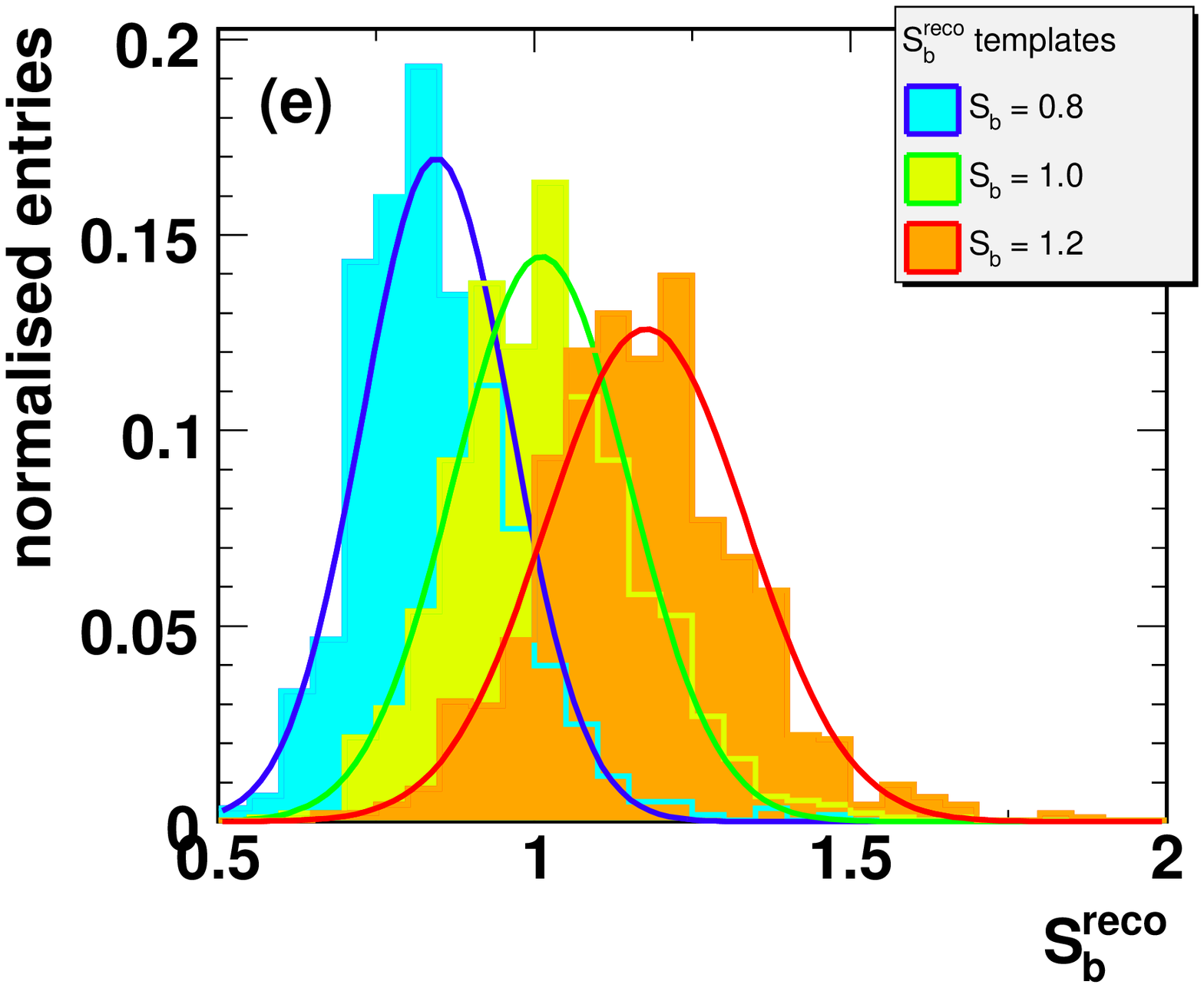}
\hspace{-0.03\textwidth}
\includegraphics[width=0.49\textwidth]{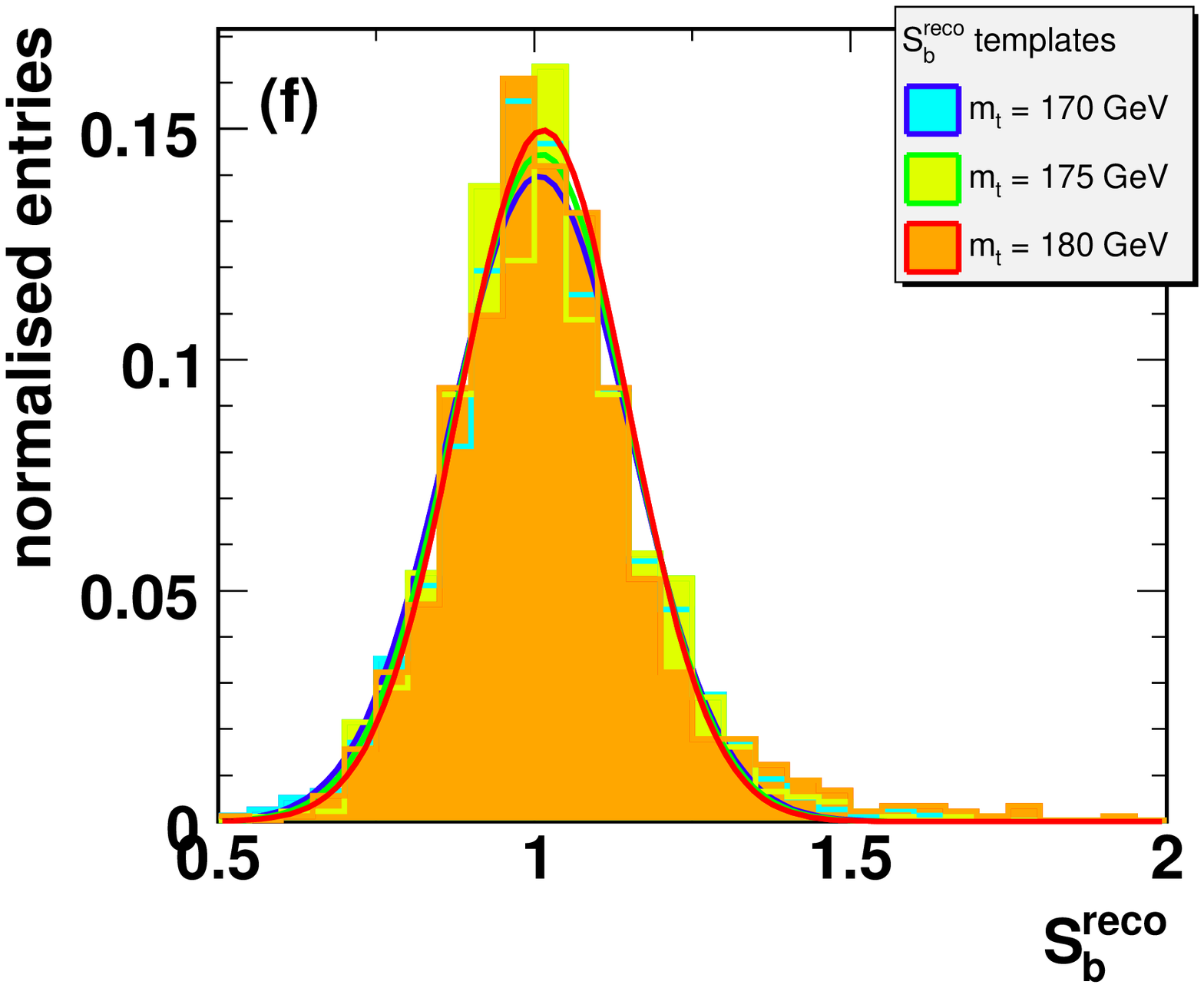}\vspace{-1ex}
\caption{\label{estimators.fig}
  Estimator distributions for simulated events with different input 
  parameters.  All distributions are normalised.
  Parameter values used in the event generation are 
  $\mtop=175\,\GeV$, $\jes=\bjes=1$, and
  $\jetsmear=1\sqrt{\GeV}$ except as noted in the plots.
  Plot (a) shows how the \mtopreco template depends 
  on the input \mtop value, plot (b) shows 
  the dependence on the \jetsmear value.
  Similar plots are shown in 
  (c) and (d) for the dependence of the \jesreco distribution on 
  the input \jes and \jetsmear values.  In (e), the dependence of
  the \bjesreco template on the \bjes input is visualized, while plot
  (f) shows that there is no large dependence of the \bjesreco
  distribution on the \mtop input value.
  The functions used to parameterize the templates as described
  in Section~\ref{toymc.sec} are overlaid.}
\end{center}
\end{figure}

Estimator distributions for various choices of input parameters
are shown in Figure~\ref{estimators.fig}.
This figure shows clearly the sensitivity to a variation of the 
input parameters.

\section{Template Fit}
\label{toymc.sec}
Estimator distributions have been determined 
from the simulated events described above for various \mtop, 
\jes, \bjes, and \jetsmear input values.
To become insensitive to statistical fluctuations of these templates,
functions $T_e$ have been derived to describe each of the three
estimator distributions, $e$\,=\,\mtopreco, \jesreco, \bjesreco.
The \mtopreco templates are described with a
normalized double Gaussian, and 
normalized single Gaussian functions are used to describe the
\jesreco and \bjesreco templates.
The parameters that describe these functions are themselves
taken to be linear functions of 
the \mtop, \jes, \bjes, and \jetsmear input values.
The fact that this simple parameterization cannot describe
every detailed aspect of the templates is no limitation
for this conceptual study.
It will lead to small deviations
between fitted and input values, see below; however, the 
general conclusion that all four quantities can be measured
independently of each other remains valid.

Pseudo-experiments are then performed using simulated events as pseudo-data.
A binned likelihood fit is performed to determine \mtop, \jes, \bjes,
and \jetsmear in each pseudo-experiment, with the likelihood ${\cal L}$ as a 
function of \mtop, \jes, \bjes, and \jetsmear hypotheses given by
\begin{equation}
  \label{likelihood.eqn}
    -\ln {\cal L} \left( \mtop,\, \jes,\, \bjes,\, \jetsmear \right)
  =
    \ \ \ - \!\!\!\!\! \mathop{\sum_{e\,=\,\mtopreco,}^{}}_{\jesreco,\,\bjesreco}
      \sum_{i=1}^{n_{\rm bins}(e)}
      D_e(i)\; \ln T_e(\mtop,\, \jes,\, \bjes,\, \jetsmear;\; i) 
  \, .
\end{equation}
The first sum is over the three estimator quantities,
generically denoted by $e$, and the second sum runs over the bins $i$
of each estimator distribution.
The number of pseudo-data events in bin $i$ of estimator distribution
$e$ is denoted by $D_e(i)$, while 
$T_e(\mtop,\, \jes,\, \bjes,\, \jetsmear;\; i)$ is the value
of the function used to parameterize the template, evaluated at the 
center of bin $i$ under the assumption of given
\mtop, \jes, \bjes, and \jetsmear values.
In each pseudo-experiment, the four-dimensional $-\ln {\cal L}$ space
is scanned, and the values of \mtop, \jes, \bjes, and \jetsmear that
minimize $-\ln {\cal L}$ are taken as measured values.

\begin{figure}[p]
\begin{center}
\includegraphics[width=0.49\textwidth]{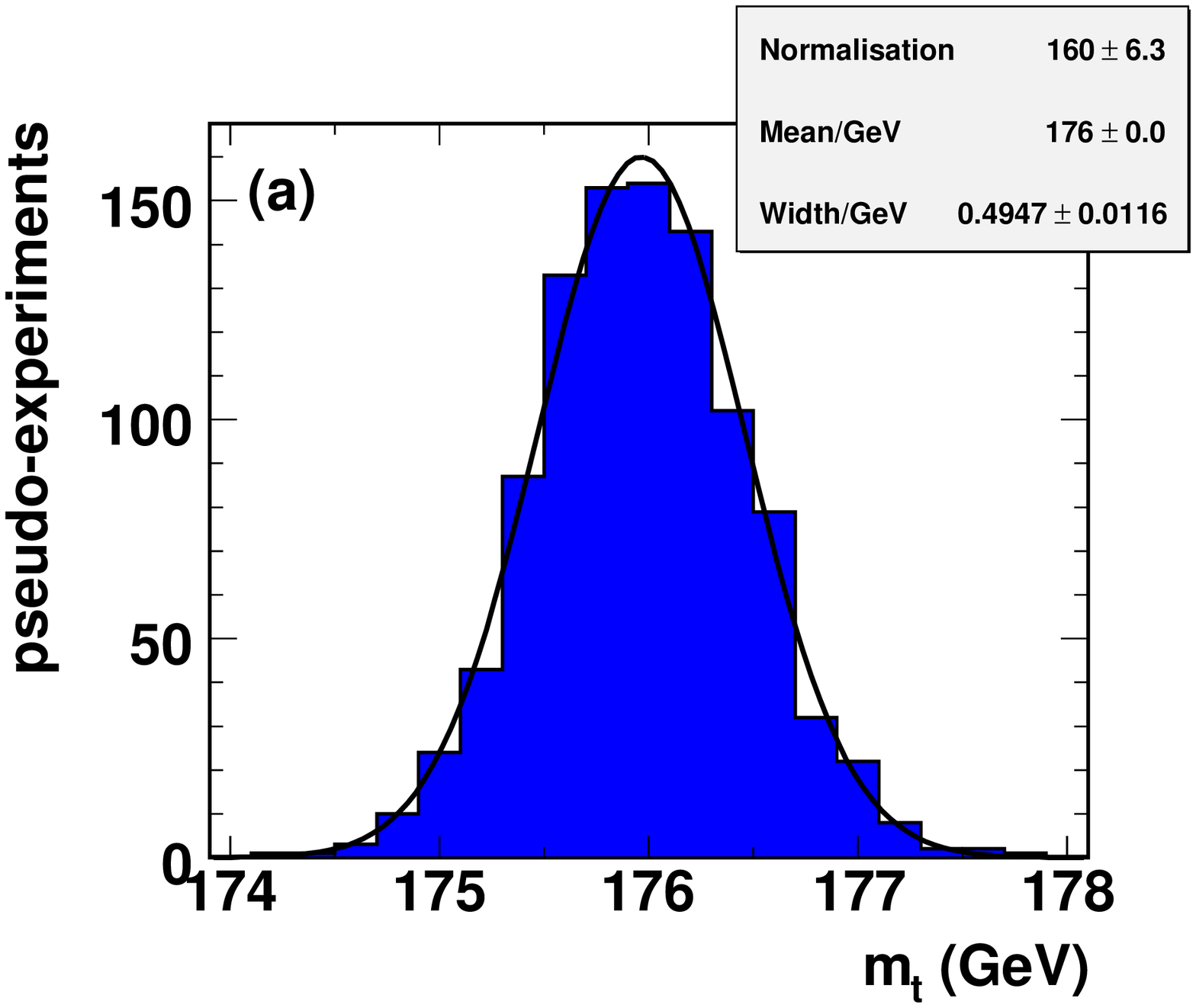}
\hspace{-0.01\textwidth}
\includegraphics[width=0.49\textwidth]{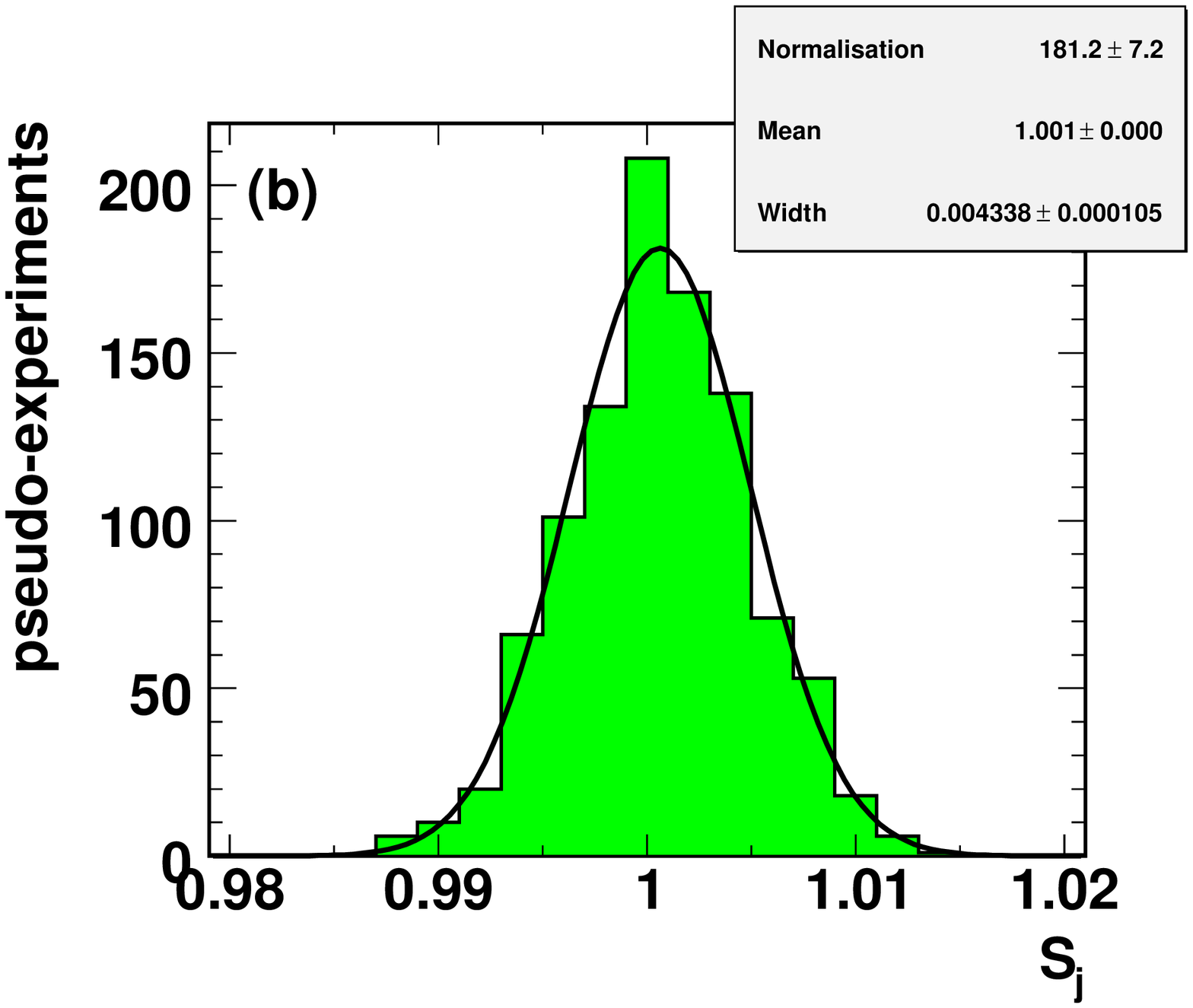}\vspace{-1ex}
\includegraphics[width=0.49\textwidth]{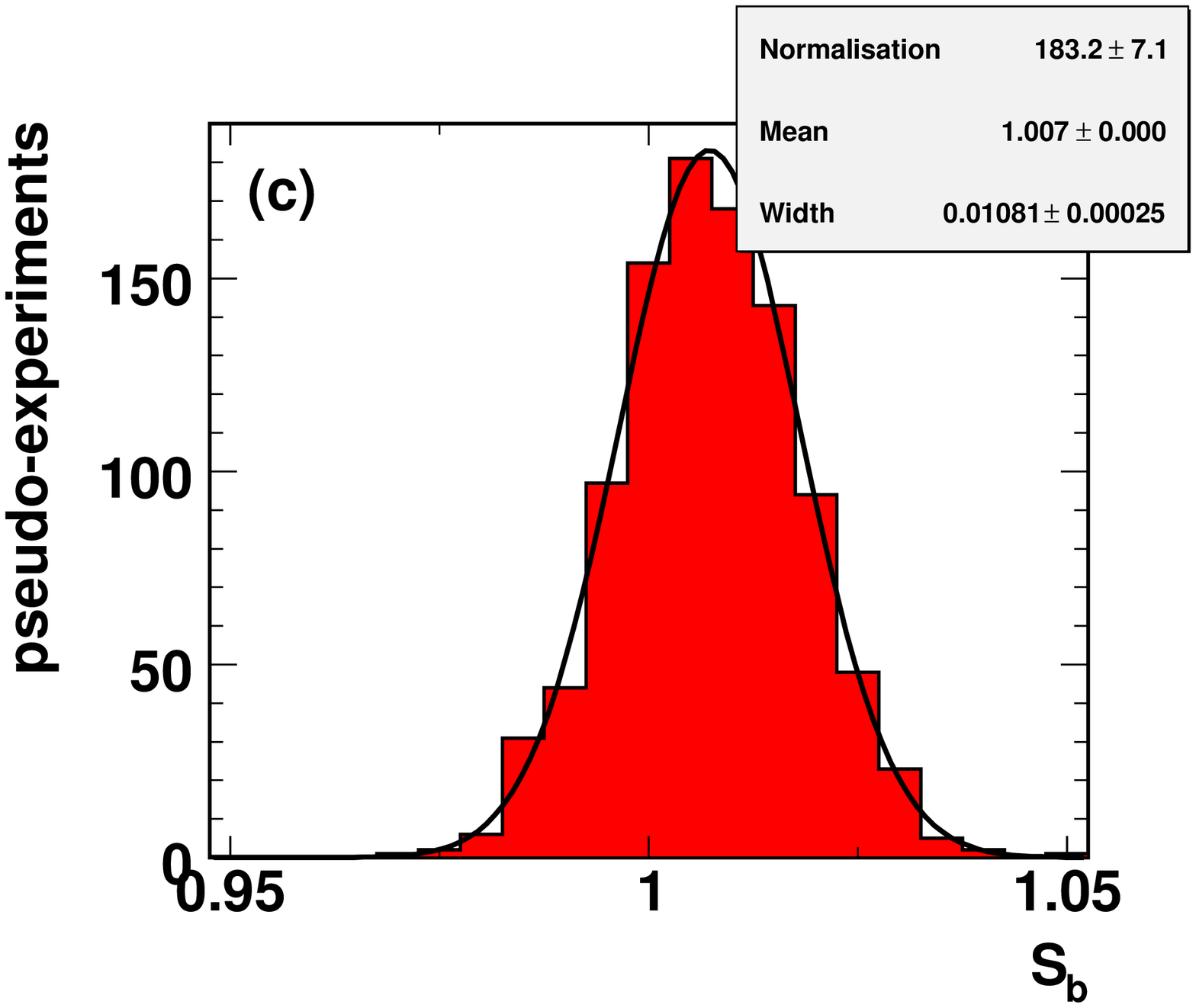}
\hspace{-0.01\textwidth}
\includegraphics[width=0.49\textwidth]{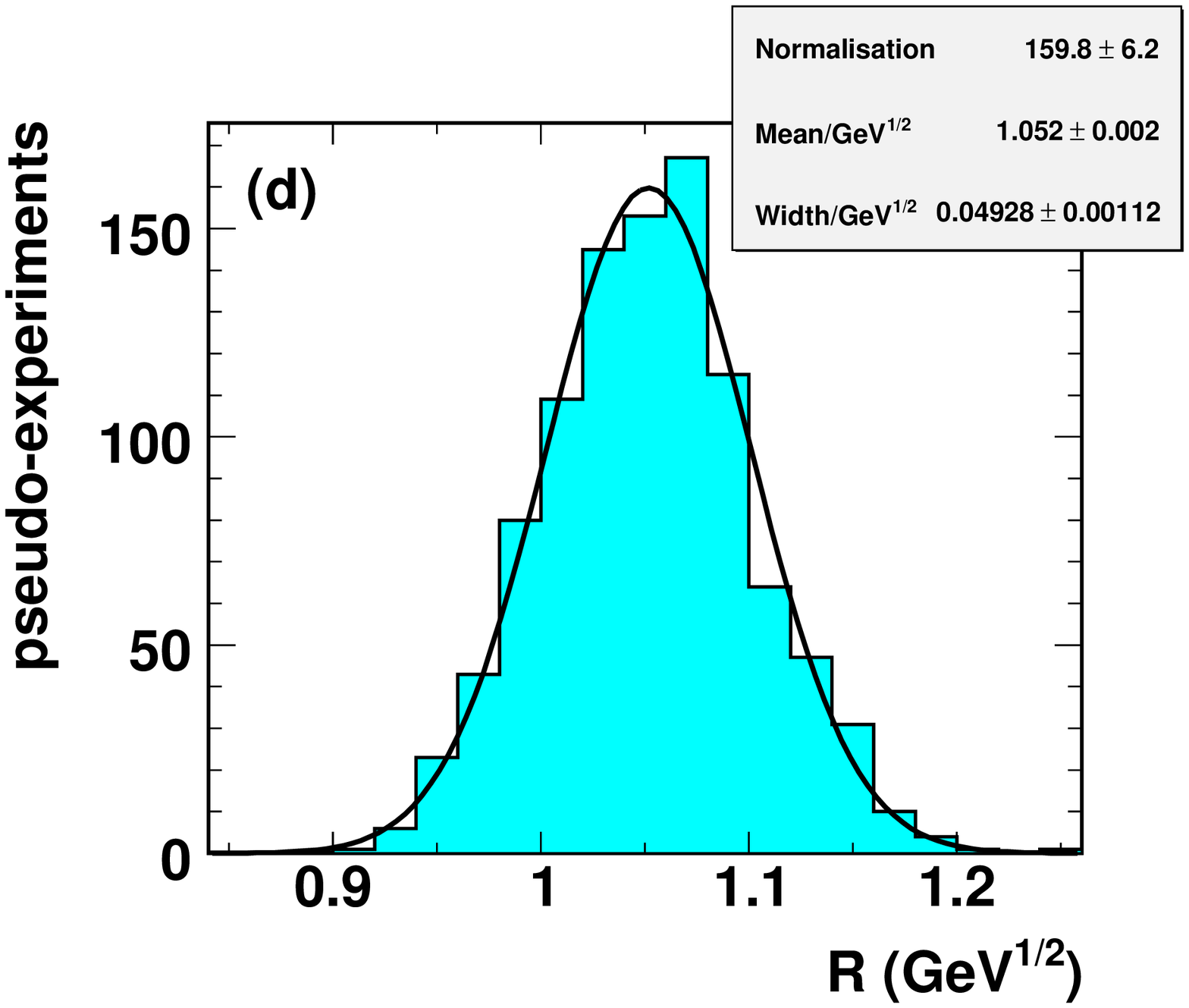}\vspace{-1ex}
\includegraphics[width=0.49\textwidth]{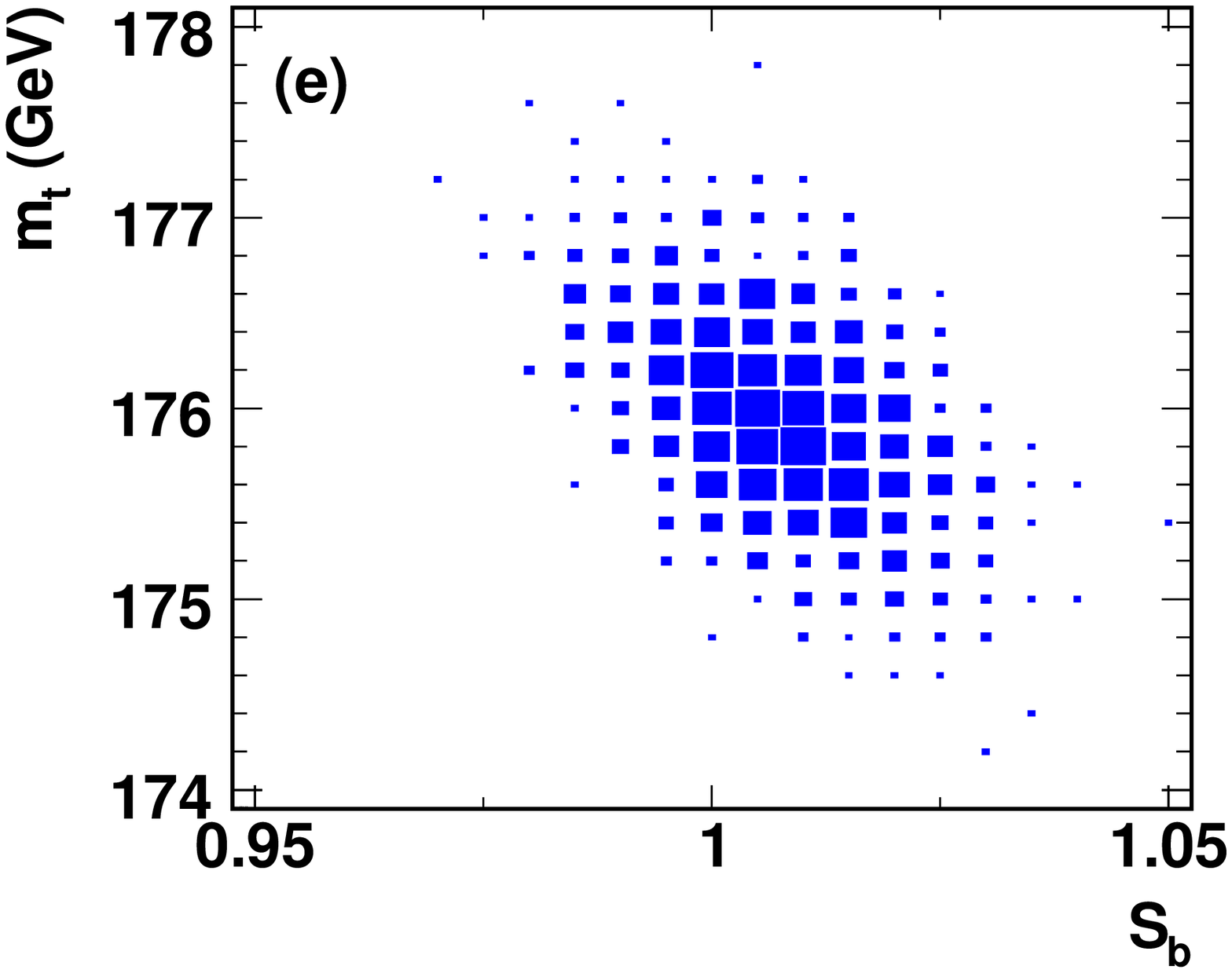}
\hspace{-0.01\textwidth}
\includegraphics[width=0.49\textwidth]{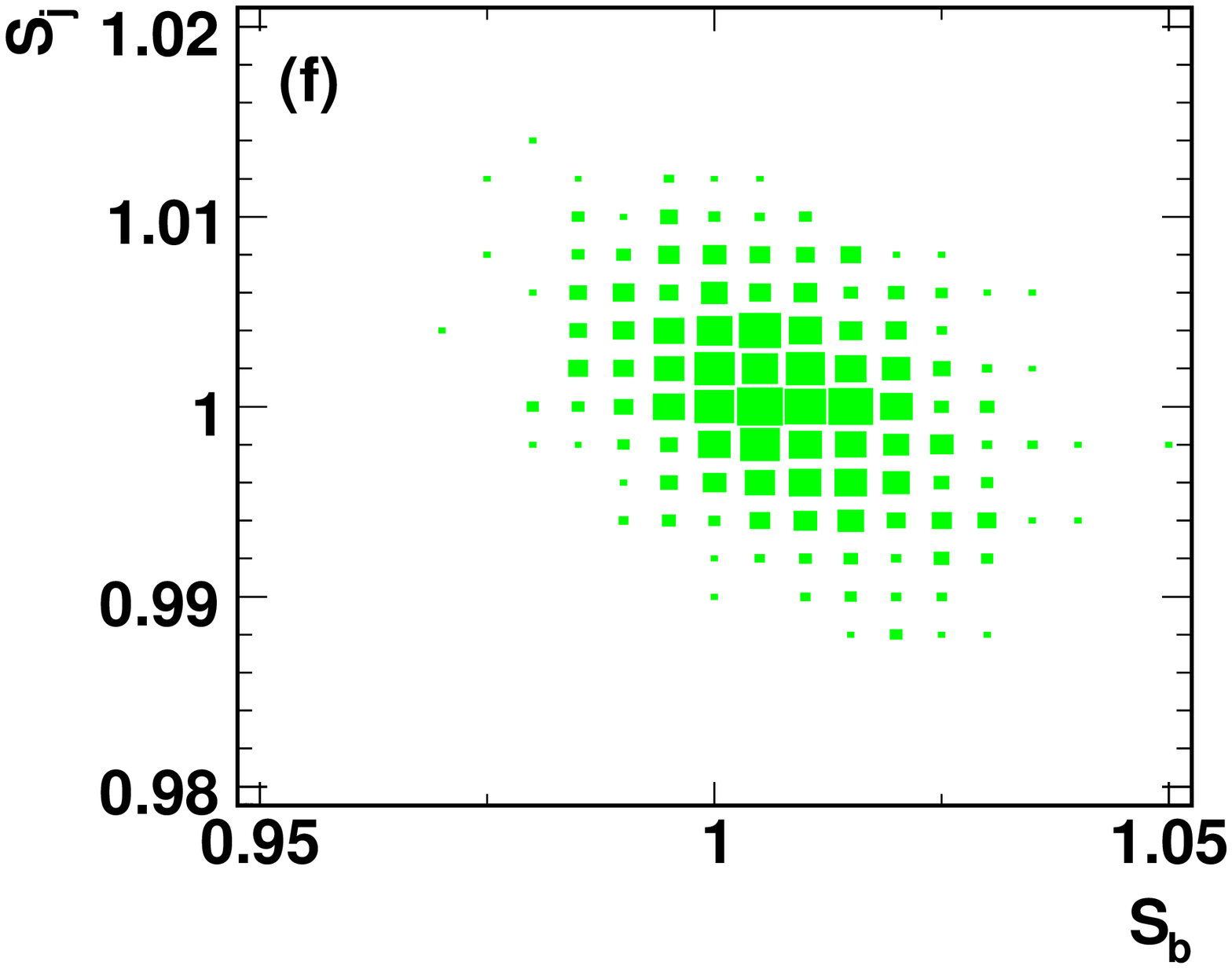}\vspace{-1ex}
\caption{\label{ensresults.fig}
  Distributions of fit results for (a) \mtop, (b) \jes, (c) \bjes,
  and (d) \jetsmear for pseudo-experiments with 300 events each, 
  generated with parameter values of 
  $\mtop=175\,\GeV$, $\jes=\bjes=1$, and
  $\jetsmear=1\sqrt{\GeV}$.
  The correlations between parameters are taken into account.
  All distributions are 
  fitted with a Gaussian; the resulting mean values and resolutions
  are given in the plots.
  The correlation between \mtop and \bjes is shown in (e), and that 
  between \jes and \bjes in (f).
}
\end{center}
\end{figure}

\begin{figure}[p]
\begin{center}
\includegraphics[width=1.0\textwidth]{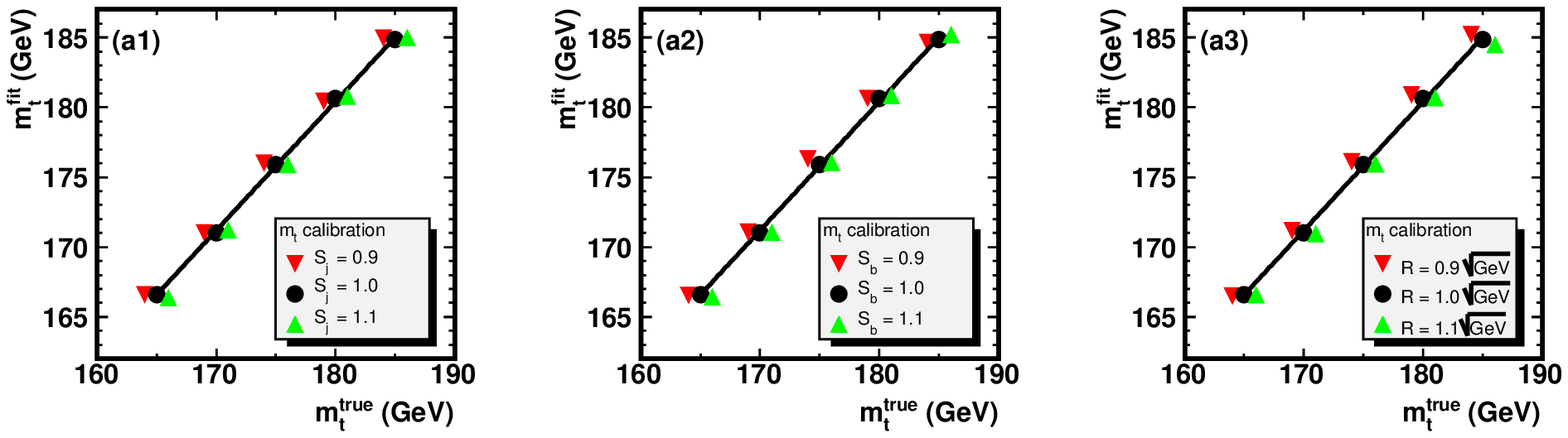}\vspace{-0.8ex}\\
\includegraphics[width=1.0\textwidth]{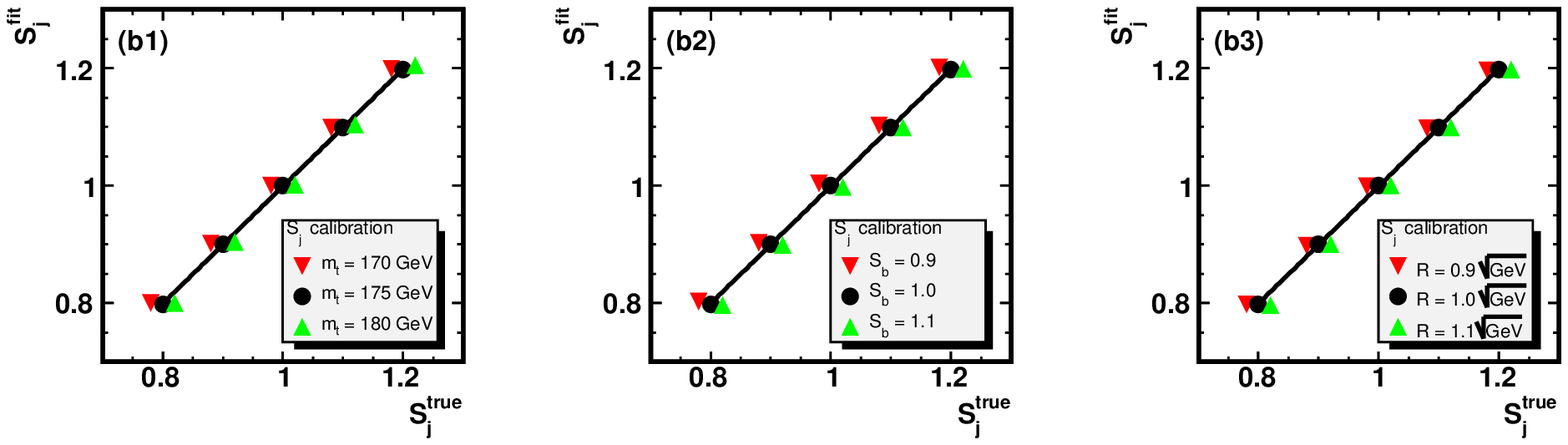}\vspace{-0.8ex}\\
\includegraphics[width=1.0\textwidth]{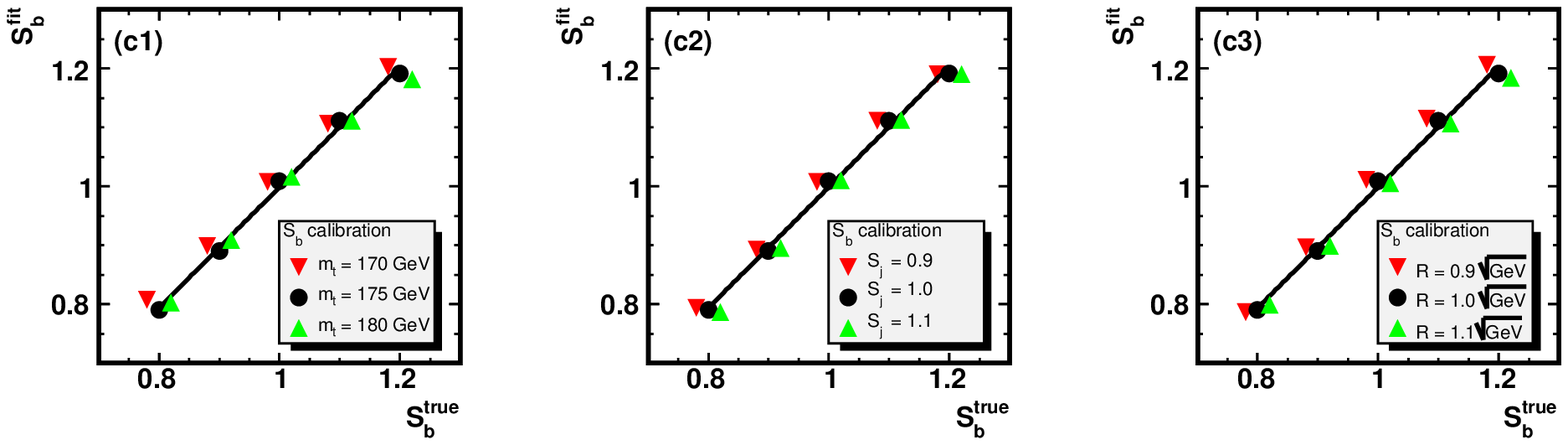}\vspace{-0.8ex}\\
\includegraphics[width=1.0\textwidth]{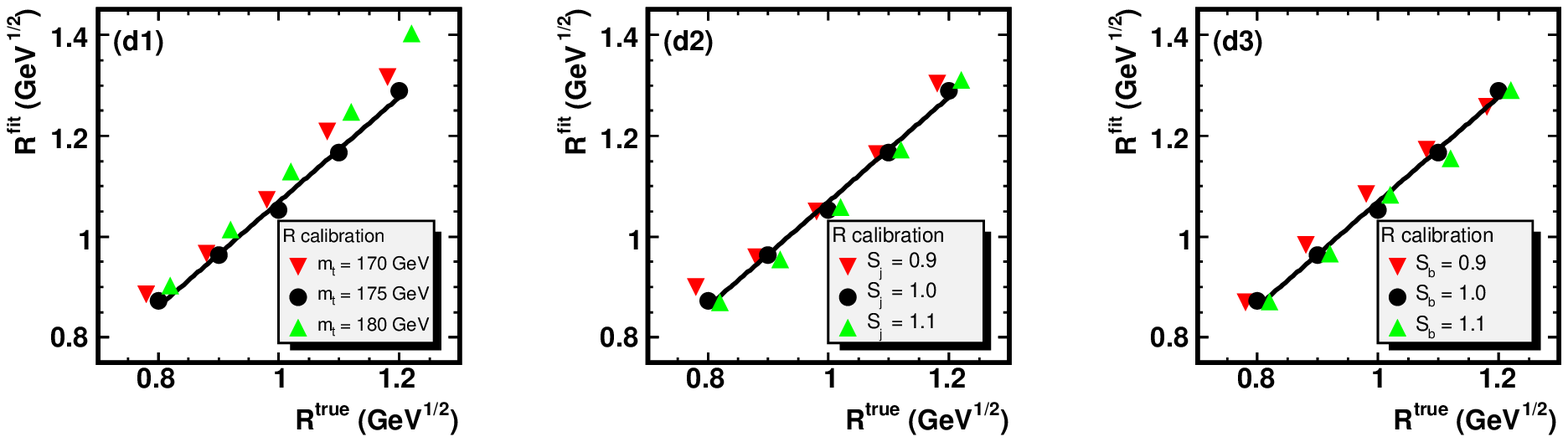}\vspace{-0.8ex}\\
\caption{\label{calibration.fig}
  Calibration curves for the determination of (a) \mtop, (b) \jes, 
  (c) \bjes, and (d) \jetsmear.  For each parameter, calibration 
  curves are shown for various input values for the other three
  parameters (left, middle, and right columns).  
  Each circle or triangle shows the mean result of pseudo-experiments
  for a given set of input values.
  The black circles are identical in all 
  three plots in a row, and the line shows the result of 
  a linear fit to these points.
  The red downward and green upward triangles show the 
  results of pseudo-experiments when 
  one of the input parameters is varied as specified in each figure.
  A slight horizontal offset has been applied to the triangles to
  render them visible.
  Parameters whose values are not noted explicitly in a plot
  are set to the default
  values of 
  $\mtop=175\,\GeV$, $\jes=\bjes=1$, and
  $\jetsmear=1\sqrt{\GeV}$.}
\end{center}
\end{figure}

For various sets of input parameter values,
1000 pseudo-experiments with 300 events each have been evaluated.
For $\mtop=175\,\GeV$, this
corresponds to an integrated luminosity of $100\,\ipb/\epsilon_b^2$
at the LHC,
where $\epsilon_b$ is the efficiency of identifying a \bquark-quark
jet.
For a given set of inputs, the distributions of measurement
values have been fitted with a Gaussian to determine the expected
measurement value and its statistical uncertainty, as shown 
in Figure~\ref{ensresults.fig}.
With an integrated luminosity of $100\,\ipb/\epsilon_b^2$, the LHC
experiments will be able to make measurements with the following statistical
uncertainties:
\begin{eqnarray}
\Delta\mtop{\rm (stat.)}  & = & 500\,\MeV \\
\nonumber
\Delta\jes{\rm (stat.)}   & = & 0.004 \\
\nonumber
\Delta\bjes{\rm (stat.)}  & = & 0.01 \\
\nonumber
\Delta\jetsmear{\rm (stat.)}  & = & 0.05\,\sqrt{\GeV} 
\,.
\end{eqnarray}
The information from the pseudo-experiments
is then used to obtain the calibration curves in
Figure~\ref{calibration.fig}, which show that all four input 
parameters can be measured independently, i.e.\ the mean measurement 
value has a large slope with respect to the input value of the 
same parameter but no dependence on the values of the other parameters.
The correlation matrix between the four measured parameters is given
in Table~\ref{correlations.table}.

The bias on the top quark mass and jet energy resolution is 
not unexpected and is due to 
the approximative description of the templates by the 
fitted functions.
Such effects could be alleviated by using more complicated 
functions, but that would go beyond the scope of this
conceptual study, and
the exact template shapes will change anyway when background and
full detector simulation are included.
In particular, it should be noted that a correction for biases of 
the raw fit results based on similar calibration curves will always
be necessary in a measurement.
The bias itself does not necessarily correspond
to a systematic uncertainty; rather, systematic errors
(due e.g.\ to uncertainties in the simulation of events) can be
determined by studying 
the {\em differences} between calibration curves obtained
with different simulation parameters.

\begin{table}[htbp]
\begin{center}
\begin{tabular}{r|cccc|}
            & \mtop & \jes & \bjes & \jetsmear \\
\hline
  \mtop     & $\phantom{-}1.0\enspace$ 
            & $-0.09$ 
            & $-0.50$
            & $-0.22$ \\
  \jes      & $-0.09$
            & $\phantom{-}1.0\enspace$ 
            & $-0.38$
            & $-0.11$ \\
  \bjes     & $-0.50$
            & $-0.38$
            & $\phantom{-}1.0\enspace$ 
            & $-0.14$ \\
  \jetsmear & $-0.22$ 
            & $-0.11$ 
            & $-0.14$ 
            & $\phantom{-}1.0\enspace$ \\
\hline
\end{tabular}
\caption{\label{correlations.table}Correlations between the four
  measured parameters for input values of $\mtop=175\,\GeV$, 
  $\jes=\bjes=1$, and $\jetsmear=1\,\sqrt{\GeV}$.}
\end{center}
\end{table}

\section{Conclusion and Outlook}
\label{outlook.sec}
While it is known how the absolute energy scale for light-quark jets
can be determined 
using events taken at hadron collider experiments (using either photon+jet
events, balancing the photon and jet in the transverse plane, or
\ljets \ttbar events, exploiting the mass constraint from the
hadronically decaying \W boson), independent information on the 
energy scale for \bquark-quark jets has so far not been obtained.
On the other hand, uncertainties on the relative difference
between the energy scales for light- and \bquark-quark jets lead to 
one of the major systematic uncertainties on measurements of the top 
quark mass at the Tevatron experiments.
In this paper, it is shown how \ljets \ttbar events can be used to 
determine the top quark mass \mtop, light-quark jet energy scale \jes, 
\bquark-quark jet energy scale \bjes, and a jet resolution parameter 
\jetsmear independently from each other.
The kinematic reconstruction of estimators on an event-by-event basis
is discussed, and results of pseudo-experiments with conditions
similar to those at the LHC experiments ATLAS and CMS are shown.
The technique presented will enable the LHC experiments 
to measure the top quark mass while at the same time 
obtaining information on the energy scale for \bquark-quark jets.
This is crucial for measurements of decays of the Higgs boson or 
of supersymmetric particles.
As increasingly large datasets are accumulated by the Tevatron 
experiments, the technique may also be used by the CDF and \dzero
experiments to reduce systematic errors on the top quark mass measurement.

The study presented here is to be viewed as a proof of principle.
In the future, the technique will be applied to fully simulated ATLAS
events, including background events.
Systematic studies need to be performed to assess how much
the results depend for example on additional gluon radiation off the 
initial- or final-state partons.
Finally, with the large event samples expected at the LHC experiments
in mind, the method has been implemented so far based on a template
fit.
The determination of the \bquark-quark jet energy scale can also
be included naturally in the Matrix Element method used at the
Tevatron for top quark mass measurements, which maximizes the
statistical sensitivity (but also needs much longer computation time
per selected event).
Any of these studies would go beyond the scope of this paper but will be 
addressed in future publications.

\section*{Acknowledgements}
The author would like to thank S.~Menke, R.~Nisius, and
J.~Schwindling for very helpful discussions.

\end{document}